\ifx\submissionver \undefined
\documentclass{article}

\newtheorem{theorem}{Theorem}[section]
\newtheorem{lemma}[theorem]{Lemma}

\usepackage{array, latexsym, graphicx,times,amsfonts,picture,color,multirow,epsf,epsfig,mathtools}
\usepackage{amsmath}
\usepackage{verbatim}
\usepackage{blkarray}
\usepackage{setspace}
\usepackage{subfigure}
\usepackage{breqn}
\usepackage{listings}
\usepackage{enumitem}
\usepackage[T1]{fontenc}
\usepackage{textcomp}
\usepackage{hyperref}
\usepackage[utf8]{inputenc}

\begin{document}
\title{Blockchain BFT Protocol for Complete Asynchronous Networks}
\author{Yongge Wang\\ UNC Charlotte}
%\date{April 3, 2020}

\maketitle

\begin{abstract}
Ethereum Research team has proposed a family of Casper blockchain consensus protocols for Ethereum 2.0. 
It has been shown in the literature that Casper Friendly Finality Gadget (Casper FFG)  for Ethereum 2.0's beacon
network cannot achieve liveness property in partially synchronous networks such as  the Internet environment.  
The ``Correct-by-Construction'' family of Casper blockchain consensus protocols (CBC Casper) 
has been proposed as a finality gadget for the future release of Ethereum 2.0 blockchain. 
Unfortunately,  neither constructive finality rule 
nor satisfactory liveness property has been obtained for CBC Casper,
and it is commonly believed that CBC Casper could not 
achieve liveness property in asynchronous networks. This paper provides 
the first probabilistic CBC Casper protocol that achieves liveness property against $t=\lfloor \frac{n-1}{3}\rfloor$
Byzantine participants in complete asynchronous networks. 
\end{abstract}

\section{Introduction}
Consensus is hard to achieve in open networks such as partial synchronous networks or complete asynchronous networks. 
Several practical protocols such as Paxos \cite{lamport1998part} and Raft \cite{ongaro2014search} have been 
designed to tolerate $\lfloor\frac{n-1}{2}\rfloor$ non-Byzantine faults. For example, Google, Microsoft, IBM, and Amazon have used Paxos in their storage or cluster management systems.
Lamport, Shostak, and Pease \cite{lamport1982byzantine} and Pease, Shostak, and Lamport \cite{pease1980reaching} 
initiated the study of reaching consensus in face of Byzantine failures  and designed the first synchronous 
solution for Byzantine agreement. For asynchronous networks, 
Fischer, Lynch, and Paterson \cite{fischer1982impossibility} showed that 
there is no deterministic protocol for the BFT problem in face of a single failure. 
Several researchers have tried to design BFT consensus protocols to circumvent the impossibility. 
The first category of efforts is to use a probabilistic approach to design BFT consensus  
protocols in completely asynchronous networks. This kind of work was initiated by 
Ben-Or \cite{ben1983another} and Rabin \cite{rabin1983randomized} and extended by others 
such as Cachin, Kursawe, and  Shoup \cite{cachin2005random}. It should be noted that though probabilistic approach
was used to design BFT protocols in asynchronous networks, some researchers 
used probabilistic approach to design BFT protocols for complete synchronous networks also.
For example, the probabilistic approach based BFT protocols 
\cite{feldman1997optimal,micali2016byzantine} employed in ALGORAND blockchain \cite{gilad2017algorand}
assumes a synchronous and complete point-to-point network. 
The second category of efforts was to design BFT consensus protocols in partial synchronous networks
which was initiated by Dwork, Lynch, and Stockmeyer \cite{dwork1988consensus}.

Ethereum foundation has tried to design a BFT finality gadget for their Proof of Stake (PoS) based Ethereum 2.0
blockchain. It has been shown in Wang \cite{yonggeBFTpaper} that their currently deployed Casper Friendly Finality Gadget 
(Casper FFG) \cite{buterin2019casper} for Ethereum 2.0 beacon network 
does not achieve liveness property in partially synchronous networks. 
Ethereum foundation has been advocating the ``Correct-by-Construction'' (CBC) family 
of Casper blockchain consensus protocols \cite{zamfir2017casper,zamfir2018introducing}
for their future release of Ethereum 2.0 blockchain. 
The CBC Casper the Friendly Ghost emphasizes the safety property. But it does not 
try to address the liveness requirement for the consensus process. Indeed, it explicitly says that \cite{zamfir2017casper}
``{\em liveness considerations are considered largely out of scope, and should be treated in future work}''.
Thus in order for CBC Casper to be deployable, a lot of work needs to be done since the Byzantine Agreement 
Problem becomes challenging only when both safety and liveness properties are required to 
be satisfied at the same time. It is simple to design BFT protocols that only satisfy one of the two requirements
(safety or liveness).
The Ethereum foundation community has made several efforts to design safety oracles for CBC Casper to help
participants to make a decision when an agreement is reached (see, e.g., \cite{caspercbcfaq}). However, this problem 
is at least as hard as coNP-complete problems. So no satisfactory solution has been proposed yet.
%(see, e.g., Monnot et al \cite{monnotsaintaditya}). https://hackingresear.ch/cbc-inspector/ 

CBC Casper has received several critiques from the community.  For example,  Ali et al \cite{muneebali}
concluded that ``{\em the definitions and proofs provided in \cite{zamfir2018introducing} 
result in neither a theoretically sound nor practically useful 
treatment of Byzantine fault-tolerance. We believe that considering correctness without liveness 
is a fundamentally wrong approach. Importantly, it remains unclear if the definition of the 
Casper protocol family provides any meaningful safety guarantees for blockchains}''. 
Though CBC Casper is not a deployable solution yet and it has several fundamental issues yet 
to be addressed, we think these critiques as in \cite{muneebali} may not be fair enough.
Indeed, CBC Casper provides an interesting framework for 
consensus protocol design. In particular, the algebraic
approach proposed by CBC Casper has certain advantages for describing Byzantine Fault Tolerance (BFT) protocols.
The analysis in this paper shows that the current formulation of CBC Casper could not achieve liveness
property. However, if one revises the CBC Casper's algebraic approach to include the concept of ``waiting''
and to enhance participant's capability to identify more malicious activities (that is, to consider general 
malicious activities in addition to equivocating activities),  then one can design
efficiently constructive liveness concepts for CBC Casper even in complete asynchronous networks. 

The structure of the paper is as follows. Section \ref{cbccaspersec} provides a brief review of the CBC Casper framework.
The author of \cite{zamfir2017casper} mentioned in several talks that CBC Casper does not guarantee
liveness in asynchronous networks. Section \ref{livenessCBCsec} presents a protocol which shows
that revised CBC Casper can indeed provide liveness property in asynchronous networks.

\section{System model and Byzantine agreement}
In this section, we describe our basic system model. For the Byzantine general problem,
there are $n$ participants and an adversary that is allowed to corrupt up to $t$ of them.
The adversary model is a static one wherein the adversary must decide whom to corrupt at the 
start of the protocol execution. 
For the network setting, we assume a complete asynchronous network of Fischer, Lynch,
and Paterson \cite{fischer1982impossibility}. That is, we make no assumptions about the relative speeds 
of processes or about the delay time in delivering a message. We also assume that processes do not have access to 
synchronized clocks, so algorithms based on time-outs cannot be used. We also assume
that the adversary has complete control of the network: he may schedule/reorder the delivery of messages as he wishes, 
and may drop or insert messages as he wishes.  However, we assume that all messages
are eventually delivered if the sender makes infinitely many trials to send the messages. 
The honest participants are completely passive: they simply follow the protocol steps 
and maintain their internal state between protocol steps. 

The computations made by the honest participants and the adversary are modeled as 
polynomial-time computations. We assume that public key cryptography
is used for message authentications. In particular, each participant should have authentic
public keys of all other participants. This means that if two participants 
$P_i$ and $P_j$ are honest and $P_j$ receives a message from $P_i$ over the network,
then this message must have been generated by $P_i$ at some prior point in time.
A Byzantine agreement protocol must satisfy the following properties:
\begin{itemize}
\item {\bf Safety}: If an honest participant decides on a value, then all other honest participants decides on the same value.
That is, it is computationally infeasible for an adversary to make two honest participants to decide on different values.
\item {\bf Liveness (termination)}: There exists a function $B(\cdot)$ such that all honest participants should
decide on a value after the protocol runs at most $B(n)$ steps. It should be noted that 
$B(n)$ could be exponential in $n$. In this case, we should further assume that 
$2^n$ is significantly smaller than $2^\kappa$ where $\kappa$ is the security parameter 
for the underlying authentication scheme. In other words, one should not be able to break the underlying
authentication scheme within $O(B(n))$ steps.
\item {\bf Non-triviality (Validity)}: If all honest participants start the protocol with the same initial value, 
then all honest participants that decide must decide on this value. 
\end{itemize}

\section{CBC Casper the Friendly Binary Consensus (FBC)}
\label{cbccaspersec}
CBC Casper has binary version and integer version. 
In this paper, we only consider Casper the Friendly Binary Consensus (FBC). 
Our discussion can be easily extended to general cases.
For the Casper FBC protocol, each participant repeatedly sends and receives messages to/from other participants. 
Based on the received messages, a participant can infer 
whether a consensus has been achieved. Assume that there are $n$ participants $P_1, \cdots, P_n$
and let $t<n$ be the Byzantine-fault-tolerance threshold.  The protocol proceeds from step to step 
(starting from step $0$)  until a consensus is reached. Specifically the step $s$ proceeds as follows:
\begin{itemize}
\item Let $\mathcal{M}_{i,s}$ be the collection of valid messages that $P_i$ has received from all participants
(including himself) from steps $0, \cdots, s-1$.
$P_i$ determines whether a consensus has been achieved. If a consensus has not been achieved yet,
$P_i$ sends the message 
\begin{equation}
\label{msgis}
m_{i,s}=\langle P_i, e_{i,s}, \mathcal{M}_{i,s}\rangle
\end{equation}
to all participants where 
$e_{i,s}$ is $P_i$'s estimated consensus value based on the received message set $\mathcal{M}_{i,s}$.
\end{itemize}
In the following, we describe how a participant $P_i$ determines whether a consensus has been achieved
and how a participant $P_i$ calculates the value $e_{i,s}$ from $\mathcal{M}_{i,s}$.

For a message $m=\langle P_i, e_{i,s}, \mathcal{M}_{i,s}\rangle$, let $J(m)=\mathcal{M}_{i,s}$.
For two messages $m_1, m_2$, we write $m_1\prec m_2$ if $m_2$ depends on $m_1$. That is, there 
is a sequence of messages $m_1', \cdots, m_v'$ such that 
$$\begin{array}{c}
m_1\in J(m_1')\\ 
m_1'\in J(m_2')\\
\cdots \\
m_v'\in J(m_2)
\end{array}$$
For a message $m$ and a message set $\mathcal{M}=\{m_1, \cdots, m_v\}$, we say that 
$m\prec \mathcal{M}$ if $m\in \mathcal{M}$ or $m\prec  m_j$ for some $j=1, \cdots, v$.
The {\em latest message} $m=L(P_i, \mathcal{M})$ by a participant $P_i$ in a message set 
$\mathcal{M}$ is a message $m\prec \mathcal{M}$ satisfying the following condition:
\begin{itemize}
\item There does not exist another message $m'\prec \mathcal{M}$ sent by participant $P_i$ with
$m\prec m'$.
\end{itemize}
It should be noted that the ``latest message'' concept is well defined for a participant $P_i$ 
if $P_i$ has not equivocated, where a participant $P_i$ equivocates if $P_i$ has sent two messages
$m_1\not=m_2$ with the properties that ``$m_1\not\prec m_2$ and $m_2\not\prec m_1$''.

For a binary value $b\in\{0,1\}$ and a message set $\mathcal{M}$, the score of a binary estimate for $b$ 
is defined as the number of non-equivocating participants $P_i$ whose latest message voted for $b$. That is,
\begin{equation}
\label{scoreEqu}
{\tt score}(b, \mathcal{M})=\sum_{L(P_i, \mathcal{M})=(P_i, b, *)}\lambda(P_i,\mathcal{M})
\end{equation}
where 
$$\lambda(P_i,\mathcal{M})=\left\{
\begin{array}{ll}
0 & \mbox{if }P_i\mbox{ equivocates in }\mathcal{M},\\
1 & \mbox{otherwise.}
\end{array}
\right.$$

\noindent 
{\bf To estimate consensus value:}
Now we are ready to define $P_i$'s estimated consensus value $e_{i,s}$ based on the received 
message set $\mathcal{M}_{i,s}$ as follows:
\begin{equation}
\label{estimatorequ}
e_{i,s}=\left\{ 
\begin{array}{ll}
0 & \mbox{if }{\tt score}(0, \mathcal{M}_{i,s}) >{\tt score}(1, \mathcal{M}_{i,s})\\
1 & \mbox{if }{\tt score}(1, \mathcal{M}_{i,s}) >{\tt score}(0, \mathcal{M}_{i,s})\\
b & \mbox{otherwise, where } b \mbox{ is coin-flip output}
\end{array}
\right.
\end{equation}

\noindent 
{\bf To infer consensus achievement:}
For a protocol execution, it is required that for all $i,s$, the number of equivocating participants 
in $\mathcal{M}_{i,s}$ is at most $t$.  A participant $P_i$ determines that a consensus has been achieved
at step $s$ with the received message set $\mathcal{M}_{i,s}$ if there exists $b\in \{0,1\}$ such that 
\begin{equation}
\label{livreq}
\forall s'>s: {\tt score}(b, \mathcal{M}_{i,s'})>{\tt score}(1-b, \mathcal{M}_{i,s'}).
\end{equation}

\section{Liveness of Revised CBC Casper FBC}
\label{livenessCBCsec}
From CBC Casper protocol description, it is clear that CBC Casper is guaranteed to be correct 
against equivocating participants. However, the ``inference rule for consensus achievement''
requires a mathematical proof that is based on infinitely many message sets $\mathcal{M}_{i,s'}$ for $s'>s$. 
This requires each participant to verify that for each potential set of $t$ Byzantine participants, 
their malicious activities will not overturn the inequality in (\ref{livreq}). 
This problem is at least co-NP hard. Thus even if the system reaches a consensus, the participants 
may not realize this fact. In order to address this challenge, Ethereum community provides 
three ``safety oracles'' (see \cite{caspercbcfaq}) to help participants to determine whether a consensus is obtained. 
The first ``adversary oracle'' simulates some protocol execution to see whether 
the current estimate will change under some Byzantine attacks. As mentioned previously, this kind 
of problem is co-NP hard and the simulation cannot be exhaustive generally.
The second ``clique oracle'' searches for the biggest clique of participant graph
to see whether there exist more than 50\% participants who agree on current estimate and 
all acknowledge the agreement. That is, for each message, the oracle checks to see if, 
and for how long, participants have seen each other agreeing on the value of that message. 
This kind of problem is equivalent to the complete bipartite graph
problem which is NP-complete. The third ``Turan oracle'' uses Turan's Theorem 
to find the minimum size of a clique that must exist in the participant edge graph. 
In a summary, currently there is no satisfactory approach for CBC Casper participants
to determine whether finality has achieved. Thus no liveness is guaranteed for CBC Casper.
Indeed, we can show that it is impossible to achieve liveness in CBC Casper.

\subsection{Impossibility of achieving liveness in CBC Casper}
In this section, we use a simple example to show that without a protocol revision, 
no liveness could be achieved in CBC Casper.
Assume that there are $3t+1$ participants. Among these participants, $t-1$ of them are malicious and never vote. 
Furthermore, assume that $t+1$ of them hold value $0$ and $t+1$ of them hold value $1$. 
Since the message delivery system is controlled by the adversary,
the adversary can let the first $t+1$ participants to receive $t+1$ voted $0$ and $t$ voted $1$. 
On the other hand, the adversary can let the next $t+1$ participants to receive $t+1$ voted 1 and $t$ voted $0$.
That is, at the end of this step, we still have that $t+1$ of them hold value $0$ and $t+1$ of them hold value $1$. 
This process can continue forever and never stop. 

In CBC Casper FBC \cite{zamfir2017casper,zamfir2018introducing}, a participant is identified 
as malicious only if he equivocates. This is not sufficient to guarantee liveness (or even safety) of the protocol.
For example, if no participant equivocates and no participant follows the equation (\ref{estimatorequ})
for consensus value estimation, then the protocol may never make a decision (that is, the protocol 
cannot achieve liveness property).  However, the protocol execution satisfies the valid protocol 
execution condition of \cite{zamfir2017casper,zamfir2018introducing} since there is zero equivocating participant.

\subsection{Revising CBC Casper FBC}
CBC Casper  does not have an in-protocol fault tolerance threshold
and does not have any timing assumptions. Thus the protocol works well in complete 
asynchronous settings. Furthermore, it does not specify when a participant $P_i$ should
broadcast his step $s$ protocol message to other participants.
That is, it does not specify when $P_i$ should stop waiting for more messages to be included $\mathcal{M}_{i,s}$.
We believe that CBC Casper authors do not specify the time for a participant to send its step $s$ protocol
messages because they try to avoid any timing assumptions. 
In fact, there is a simple algebraic approach to specify this without 
timing assumptions.  First, we revise the message set $\mathcal{M}_{i,s}$ as the collection of 
messages that $P_i$ receives from all participants (including himself) during step $s-1$. That is,
the message set $\mathcal{M}_{i,s}$ is a subset of $E_s$ where $E_s$  is defined recursively as follows:
$$\begin{array}{l}
E_0=\emptyset\\
E_1=\{\langle P_j, b, \emptyset \rangle: j=1,\cdots, n; b=0,1\} \\
E_2=\{\langle P_j, b, \mathcal{M}_{j,1} \rangle: j=1,\cdots, n; b=0,1; \mathcal{M}_{j,1} \subset E_1\}  \\
\cdots\\
E_s=\{\langle P_j, b, \mathcal{M}_{j,s-1} \rangle: j=1,\cdots, n; b=0,1; \mathcal{M}_{j,s-1} \subset E_{s-1}\}  \\
\cdots 
\end{array}$$
Then we need to revise the latest message definition $L(P_j,\mathcal{M}_{i,s})$ accordingly:
\begin{equation}
\label{latsmsg}
L(P_j,\mathcal{M}_{i,s})=\left\{\begin{array}{ll}
m&\mbox{if }\langle P_j, b, m\rangle\in \mathcal{M}_{i,s}\\
\emptyset&\mbox{otherwise}
\end{array}\right.
\end{equation}
As we have mentioned in the preceding section, CBC Casper FBC \cite{zamfir2017casper,zamfir2018introducing}
only considers equivocating as malicious activities. This is not sufficient to guarantee protocol liveness against 
Byzantine faults. In our following revised CBC Casper model, we consider any participant that does
not follow the protocol as malicious and exclude their messages:
\begin{itemize}
\item For a message set $\mathcal{M}_{i,s}$, let $I(\mathcal{M}_{i,s})$ be the set of identified malicious participants
from $\mathcal{M}_{i,s}$. Specifically, let 
$$I(\mathcal{M}_{i,s})=E(\mathcal{M}_{i,s})\cup F(\mathcal{M}_{i,s})$$
where $E(\mathcal{M}_{i,s})$ is the set of equivocating participants within $\mathcal{M}_{i,s}$
and $F(\mathcal{M}_{i,s})$ is the set of participants that does not follow the protocols within $\mathcal{M}_{i,s}$.
For example, $F(\mathcal{M}_{i,s})$ includes participants that do not follow the consensus value estimation process
properly or do not wait for enough messages before posting his own protocol messages.
\end{itemize} 
With the definition of $I(\mathcal{M}_{i,s})$, we should also redefine the score function (\ref{scoreEqu})
by revising the definition of $\lambda(P_i,\mathcal{M})$ accordingly:
$$\lambda(P_i,\mathcal{M})=\left\{
\begin{array}{ll}
0 & \mbox{if }P_i\in I(\mathcal{M}),\\
1 & \mbox{otherwise.}
\end{array}
\right.$$

\subsection{Secure BFT protocol in the revised CBC Casper}
\label{cbcbftsec}
With the revised CBC Casper, we are ready to introduce the ``waiting'' concept and 
specify when a participant $P_i$ should send his step $s$ protocol message:
\begin{itemize}
\item A participant $P_i$ should wait for at least $n-t+|I(\mathcal{M}_{i,s})|$ valid messages $m_{j,s-1}$ from
other participants before he can broadcast his step $s$ message $m_{i,s}$.  That is, $P_i$ should wait 
until $|\mathcal{M}_{i,s}|\ge n-t+|I(\mathcal{M}_{i,s})|$ to broadcast his step $s$ protocol message.
\item In case that a participant $P_i$ receives $n-t+|I(\mathcal{M}_{i,s})|$ valid messages $m_{j,s-1}$ from
other participants (that is, he is ready to send step $s$ protocol message) before he could post his
step $s-1$ message, he should wait until he finishes sending his step $s-1$ message.
\item After a participant $P_i$ posts his step $s$ protocol message, it should discard 
all messages from steps $s-1$ or early except decision messages that we will describe later.
%it will still record protocol messages 
%received from other participants for steps $s-1$ or less for consensus finality purpose.
\end{itemize}
It is clear that these specifications does not have any restriction on the timings. Thus the protocol works
in complete asynchronous networks.   

In Ben-Or's BFT protocol \cite{ben1983another}, if consensus is not achieved yet, 
the participants autonomously toss a coin until more than $\frac{n+t}{2}$ participant outcomes coincide. 
For Ben-Or's maximal Byzantine fault tolerance threshold
$t\le\lfloor \frac{n}{5}\rfloor$, it takes exponential steps of coin-flipping to converge.
It is noted that, for $t=O(\sqrt{n})$, Ben-Or's protocol takes constant rounds to converge.
Bracha \cite{bracha1984asynchronous} improved Ben-Or's protocol to defeat $t<\frac{n}{3}$ Byzantine faults.
Bracha first designed a reliable broadcast protocol with the following properties (Bracha's reliable broadcast protocol
is briefly reviewed in the Appendix):
If an honest participant broadcasts a message, then all honest participants will receive the same
message in the end. If a dishonest participants $P_i$ broadcasts a message, then either all honest 
participants accept the identical message or no honest participant accepts any value from $P_i$.
By using the reliable broadcast primitive and other validation primitives, 
Byzantine participants are transformed to fail-stop participants in Bracha \cite{bracha1984asynchronous}.
In this section, we assume that a reliable broadcast primitive such as the one by Bracha's is used
in our protocol execution.
In the following, we adapt Bracha's BFT protocol to the CBC Casper framework.
At the start of the protocol, each participant $P_i$ holds an initial value 
in his variable $x_i\in\{0,1\}$. The protocol proceeds from step to step.
The step $s$ consists of the following sub-steps.
\begin{enumerate}
\item Each participant $P_i$ reliably broadcasts $\langle P_i, x_i, \mathcal{M}_{i,s,0}\rangle$ to 
all participants where $\mathcal{M}_{i,s,0}$ is the message set that $P_i$ has received during step $s-1$.
Then $P_i$ waits until it receives $n-t$ valid messages in $\mathcal{M}_{i,s,1}$ and 
computes the estimate $e_{i,s}$ using the value estimation function (\ref{estimatorequ}).
\item Each participant $P_i$ reliably  broadcasts $\langle P_i, e_{i,s}, \mathcal{M}_{i,s,1}\rangle$ to 
all participants and waits until it receives $n-t$ valid messages in $\mathcal{M}_{i,s,2}$.
If there is a $b$ such that ${\tt score}(b, \mathcal{M}_{i,s,2})>\frac{n}{2}$, then let 
$e_{i,s}'=b$ otherwise, let $e_{i,s}'=\perp$.
\item Each participant $P_i$ reliably  broadcasts $\langle P_i, e_{i,s}', \mathcal{M}_{i,s,2}\rangle$ to 
all participants and waits until it receives $n-t$ valid messages in $\mathcal{M}_{i,s,3}$.
$P_i$ distinguishes the following three cases:
\begin{itemize}
\item If ${\tt score}(b, \mathcal{M}_{i,s,2})>2t+1$ for some $b\in\{0,1\}$,
then $P_i$ decides on $b$ and broadcasts his decision together with justification to all participants.
\item If ${\tt score}(b, \mathcal{M}_{i,s,2})>t+1$ for some $b\in\{0,1\}$,
then $P_i$ lets $x_i=b$ and moves to step $s+1$.
\item Otherwise, $P_i$ flips a coin and let $x_i$ to be coin-flip outcome. $P_i$ moves to step $s+1$.
\end{itemize}
\end{enumerate} 

Assume that $n=3t+1$. The security of the above protocol can be proved be establishing a sequence of lemmas.

\begin{lemma}
If all honest participants hold the same initial value $b$ at the start of the protocol,
then every participant decides on $b$ at the end of step $s=0$.
\end{lemma}

\noindent 
{\em Proof}. 
At sub-step 1, each honest participant receives at least $t+1$ value $b$ among the $2t+1$ received values.
Thus all honest participants broadcast $b$ at sub-step 2. If a malicious participant $P_j$ broadcasts $1-b$
during sub-step 2, then it cannot be justified since $P_j$ could not receive $t+1$ 
messages for $1-b$ during sub-step 1. Thus $P_j$ will be included in $I(\cal M)$. 
That is, each honest participant receives $2t+1$ messages for $b$ at the end of sub-step 2
and broadcasts $b$ during sub-step 3. Based on the same argument, all honest participants decide 
on $b$ at the end of sub-step 3.
\hfill$\Box$

\begin{lemma}
If an honest participant $P_i$ decides on a value $b$ at the end of step $s$, then all honest 
participants either decide on $b$ at the end of step $s$ or at the end of step $s+1$.
\end{lemma}

\noindent 
{\em Proof}. If an honest participant $P_i$ decides on a value $b$ at the end of sub-step 3,  then 
$P_i$ receives $2t+1$ valid messages for the value $b$. Since the underlying broadcast protocol
is reliable, each honest participant receives at least $t+1$ these valid messages for the value $b$.
Thus if a participant $P_i$ does not decide on the value $b$ at the end of sub-step 3, it 
would set $x_i=b$. That is, all honest participants will decide during step $s+1$.
\hfill$\Box$

\vskip 5pt
The above two Lemmas show that the protocol is a secure Byzantine Fault Tolerance protocol
against $\lfloor \frac{n-1}{3}\rfloor$ Byzantine faults in complete asynchronous networks.
The above BFT protocol may take exponentially many steps to converge. 
However, if a common coin such as the one in Rabin \cite{rabin1983randomized} 
is used, then the above protocol converges in constant steps. It should be noted that 
Ethereum 2.0 provides a random beacon which could be used as the common coin
for the above BFT protocol. Thus the above BFT protocol could be implemented with constant steps on 
Ethereum 2.0.
% If common coins could be implemented,
%one may also implement Cachin-Kursawe-Shoup protocol \cite{cachin2005random} 

\bibliographystyle{plain}

\appendix
\section{Bracha's broadcast primitive}
\label{brabroadsec}
Assume $n>3t$. Bracha \cite{bracha1984asynchronous} designed 
a broadcast protocol for asynchronous networks with the following properties:
\begin{itemize}
\item If an honest participant broadcasts a message, then all honest participants accept the message.
\item If a dishonest participant $P_i$ broadcasts a message, then either all honest 
participants accept the same message or no honest participant accepts any value from $P_i$.
\end{itemize}
Bracha's broadcast primitive runs as follows:
\begin{enumerate}
\item The transmitter $P_i$ sends the value $\langle P_i, initial, v\rangle$ to all participants.
\item If a participant $P_j$ receives a value $v$ with one of the following messages
\begin{itemize}
\item $\langle P_i, {\tt initial}, v\rangle$
\item $\frac{n+t}{2}$ messages of the type $\langle {\tt echo}, P_i, v\rangle$
\item $t+1$ message of the type $\langle {\tt ready}, P_i, v\rangle$
\end{itemize}
then $P_j$ sends the message $\langle {\tt echo}, P_i, v\rangle$ to all participants.
\item If a participant $P_j$ receives a value $v$ with one of the following messages
\begin{itemize}
\item $\frac{n+t}{2}$ messages of the type $\langle {\tt echo}, P_i, v\rangle$
\item $t+1$ message of the type $\langle {\tt ready}, P_i, v\rangle$
\end{itemize}
then $P_j$ sends the message $\langle {\tt ready}, P_i, v\rangle$ to all participants.
\item If a participant $P_j$ receives $2t+1$ messages of the type $\langle {\tt ready}, P_i, v\rangle$,
then $P_j$ accepts the message $v$ from $P_i$.
\end{enumerate}
Assume that $n=3t+1$. The intuition for the security of Bracha's broadcast primitive is as follows.
First, if an honest participant $P_i$ sends the value $\langle P_i, initial, v\rangle$, then 
all honest participant will receive this message and echo the message $v$. Then all 
honest participants send the ready message for $v$ and all honest participants accept the message $v$.

Secondly, if honest participants $P_{j_1}$  and $P_{j_2}$ send ready messages for 
$u$ and $v$ respectively, then we must have $u=v$. This is due to the following fact.
A participant $P_j$ sends a $\langle {\tt ready}, P_j, u\rangle$ message only if it receives $t+1$ ready messages
or $2t+1$ echo messages. That is, there must be an honest participant who received $2t+1$ echo messages for $u$.
Since an honest participant can only send one message of each type, this means that all honest participants
will only sends ready message for the value $u$.

In order for an honest participant $P_j$ to accept a message $u$, it must receive $2t+1$ ready messages.
Among these messages, at least $t+1$ ready messages are from honest participants. 
An honest participant can only send one message of each type. Thus if honest
participants $P_{j_1}$  and $P_{j_2}$ accept messages $u$ and $v$ respectively, then we must have 
$u=v$. Furthermore, if a participant $P_j$ accepts a message $u$, we just showed that at least 
$t+1$ honest participants have sent the ready message for $u$. In other words, all honest 
participants will receive and send at least $t+1$ ready message for $u$. By the argument 
from the preceding paragraph, each honest participant sends one ready message for $u$.  
That is, all honest participants will accept the message $u$.

\end{document}